\documentclass[letterpaper, 10 pt, conference]{ieeeconf}  

\IEEEoverridecommandlockouts                              

\usepackage{graphicx} 
\usepackage{amsmath} 
\usepackage{amssymb}  
\usepackage{subfiles}
\usepackage{subfigure}
\usepackage{cite}
\usepackage{siunitx}
\usepackage{xcolor}

\title{\LARGE \bf
Star-Tracker-Constrained Attitude MPC for CubeSats
}
\def\BibTeX{{\rm B\kern-.05em{\sc i\kern-.025em b}\kern-.08em
    T\kern-.1667em\lower.7ex\hbox{E}\kern-.125emX}}

\author{Dominik Beňo$^{1,\ast}$\thanks{$^\ast$ Both authors, Dominik Beňo and Patrik Valábek, contributed equally to this work.}, Patrik Valábek$^{2,\ast}$, Martin Hromčík$^{1}$ and Martin Klaučo$^{1}$%
\thanks{$^{1}$Dominik Beňo, Martin Hromčík and Martin Klaučo are with the Faculty of Electrical Engineering, Czech Technical University in Prague, Czech Republic.}%
\thanks{$^{2}$Patrik Valábek is with the Slovak University of Technology in Bratislava, Slovakia.}%
}

\begin{document}

\maketitle
\thispagestyle{empty}
\pagestyle{empty}

\begin{abstract}

This paper presents an online linear model predictive control (MPC) framework for slew maneuvers that maintains star-tracker availability during ground-target tracking. The nonlinear rigid-body dynamics and geometric exclusion constraints are analytically linearized about the current state estimate at each control step, yielding a time-varying linear MPC formulation cast as a standard quadratic program (QP). This structure is compatible with established aerospace flight-software practices and offers a computational profile with lower online complexity than comparable nonlinear MPC schemes. The controller incorporates angular-rate, actuator, and star-tracker exclusion constraints over a receding horizon. Performance is assessed in high-fidelity nonlinear model-in-the-loop simulations using NASA's "42" spacecraft dynamics simulator, including a Monte Carlo campaign over varying target geometries and inertia perturbations.

\end{abstract}

\section{Introduction}
\subsection{Motivation}
Low-cost agile spacecraft often operate with limited onboard computation and little sensor redundancy in the Attitude and Orbit Control System (AOCS). In ground-track imaging missions, the spacecraft must continuously reorient to follow a moving target while maintaining payload line-of-sight quality. This applies to optical imagers, push-broom instruments, synthetic aperture radar (SAR), and optical communication systems, where data quality depends on both final attitude error and transient pointing behavior during the maneuver. On many small and medium platforms, attitude determination relies on a single star-tracker because of mass, power, and cost limits. As a result, star-tracker rate and exclusion limits become hard constraints on every retargeting maneuver. The sensor availability is a control constraint rather than an implementation detail.

This dependence introduces an inherent rate constraint on every reorientation maneuver. Star-trackers cannot deliver reliable attitude measurements when the angular rate of the instrument exceeds a sensor-specific threshold~\cite{cite:fundamentals}. Mission-level agility is directly bound by this limit rather than by actuator capability alone. Star-tracker slew rate limits typically range from $1^\circ\,s^{-1}$ to $3^\circ\,s^{-1}$, depending on sensor generation and operating mode \cite{str_testing}. That is often lower than the maximum angular velocity achievable by actuators on the same satellite. Furthermore, star-tracker blinding arises from the violation of exclusion constraints imposed by bright extended bodies, most notably Earth's limb and the Sun. Operating inside both boundaries simultaneously is therefore a necessary condition for continuous, high-quality attitude determination.

A further consideration is that reacquisition of a star-tracker fix after loss of lock is uncertain in time and may require a non-negligible recovery interval \cite{str_testing, agile-str}. For a mission with a tightly scheduled ground-tracking sequence, such delays reduce useful observation time and may compromise data quality entirely. Avoiding loss of lock is therefore a primary control objective.

Several solution approaches have been proposed for maneuvers subject to these constraints. The most widely deployed in operational missions is offline guidance. Constraint-satisfying attitude trajectories are computed on the ground, validated, and uploaded to the spacecraft as reference sequences~\cite{cite:lunar_orbiter,cite:att_traj_opt}. This approach is deterministic and well supported by flight heritage, but it relies on operator intervention and a ground-to-satellite uplink cycle for every new maneuver. This limits responsiveness to unplanned targets and reduces operational autonomy~\cite{autonomy}. Artificial potential field methods steer the trajectory away from exclusion zones through repulsive potentials, but without a formal guarantee of constraint satisfaction over a finite horizon~\cite{cite:potentials_first_publication,cite:potentials}. Nonlinear model predictive control can encode the full constraint set and nonlinear dynamics simultaneously. Fast NMPC solvers with exclusion zone handling via penalty functions have been proposed \cite{att_nmpc}, but penalty-based enforcement provides only soft constraint satisfaction, and the iterative nonlinear program still introduces timing variability that is difficult to certify for fixed-cycle onboard software. Moreover, existing NMPC formulations have targeted rest-to-rest reorientation rather than continuous ground-target tracking with time-varying geometry.

None of the reviewed approaches jointly enforces pointing, star-tracker exclusion, angular rate, and actuator constraints in an online feedback loop for continuous target tracking. This paper addresses that gap.

\subsection{Contribution}
Agile ground tracking under star-tracker availability constraints remains an open problem for small and medium spacecraft. Single-star-tracker configurations impose a coupled pointing and rate constraint on every reorientation maneuver. Recent work confirms this is a binding limitation in both Earth observation \cite{hypso-mission} and laser communication missions \cite{cite:laser_pointing}.
The main contributions of this paper are:
\begin{itemize}
    \item A time-varying linear MPC formulation that jointly enforces angular-rate, actuator, and star-tracker exclusion constraints over a receding horizon, enabling online constraint-aware ground tracking without offline trajectory design or operator intervention.
    \item Analytical linearization of the nonlinear rigid-body dynamics and geometric constraint outputs about the current state estimate at each control step, reducing the online problem to a standard QP.
    \item A Monte Carlo characterization of the method's operational envelope over varying target geometries and inertia perturbations, identifying the conditions under which the successive linearization preserves constraint satisfaction and providing concrete reliability bounds for mission design.
\end{itemize}

\section{Problem statement}
\label{sec:problem_statement}

Consider a spacecraft in Low Earth Orbit that must continuously reorient to align its instrument boresight to a ground target fixed in the terrestrial frame \cite{hypso-mission}, while preserving star-tracker operability throughout the maneuver. With a single star-tracker, attitude knowledge is contingent on the sensor remaining within its operational envelope. This couples the pointing objective with constraints over the entire maneuver: minimum angular separations of the star-tracker boresight from the Sun and nadir, body angular rate within the star-tracker's tracking capability, and actuator torque limits. The control objective is to minimize pointing error while satisfying all constraints simultaneously, without relying on precomputed trajectories or ground intervention.

\subsection{Spacecraft model}
An 8U CubeSat rigid-body model in body frame $\mathcal{B}$, with the Earth-Centered Inertial J2000 frame $\mathcal{N}$ as the inertial reference~\cite{cite:fundamentals}, with state
$\boldsymbol{x}=(\boldsymbol{\omega}^\intercal,\boldsymbol{q}^\intercal)^\intercal$, where
$\boldsymbol{\omega}\in\mathbb{R}^3$ is the angular-rate state and
$\boldsymbol{q}\in\mathbb{R}^4$ is the unit, scalar-first quaternion state, is considered.
The control input is body torque $\boldsymbol{u}\in\mathbb{R}^3$. The term $\boldsymbol{L}_{\mathrm{ext}}$ embeds disturbance torques and the net effects of all moving parts within the spacecraft.
Throughout the paper, $\boldsymbol{I}_n$ denotes the $n\times n$ identity matrix and $\boldsymbol{0}_{m\times n}$ denotes the $m\times n$ zero matrix.

The nonlinear model is then given by
\begin{equation}
\dot{\boldsymbol{\omega}}=\boldsymbol{J}_{\mathcal{B}}^{-1}\left(\boldsymbol{u}+\boldsymbol{L}_{\mathrm{ext}}-\boldsymbol{\omega}\times\left(\boldsymbol{J}_{\mathcal{B}}\boldsymbol{\omega}\right)\right),
\end{equation}
\begin{equation}
\dot{\boldsymbol{q}}=\frac{1}{2}\boldsymbol{Q}(\boldsymbol{q})
\begin{pmatrix}
0 \\
\boldsymbol{\omega}
\end{pmatrix},
\end{equation}
where $\boldsymbol{J}_{\mathcal{B}}\in\mathbb{R}^{3\times 3}$ is the (non-diagonal) inertia tensor. The control input is $\boldsymbol{u}$ and $\boldsymbol{L}_\mathrm{ext}$ represents disturbance torques. The quaternion kinematics are expressed using the matrix $\boldsymbol{Q}(\boldsymbol{q})$:
\begin{equation}
\boldsymbol{Q}(\boldsymbol{q})=
\begin{pmatrix}
q_0 & -q_1 & -q_2 & -q_3 \\
q_1 &  q_0 & -q_3 &  q_2 \\
q_2 &  q_3 &  q_0 & -q_1 \\
q_3 & -q_2 &  q_1 &  q_0
\end{pmatrix}.
\end{equation}

To express constraints and error metrics, a mapping between the body and inertial frames is required. The fixed body-frame boresight vectors for instrument and star-tracker $\boldsymbol{v}_{\mathrm{ins}}^{\mathcal{B}}$ and $\boldsymbol{v}_{\mathrm{str}}^{\mathcal{B}}$, are related to inertial-frame unit vectors $\boldsymbol{v}_{\mathrm{ins}}^{\mathcal{N}}$, $\boldsymbol{v}_{\mathrm{str}}^{\mathcal{N}}$ via the direction cosine matrix (DCM) $\boldsymbol{C}(\boldsymbol{q}):\mathbb{R}^4\to\mathbb{R}^{3\times 3}$, $\boldsymbol{v}^{\mathcal{B}} = \boldsymbol{C}(\boldsymbol{q})\,\boldsymbol{v}^{\mathcal{N}}$, where for $\boldsymbol{q}=(q_0,\boldsymbol{q}_v^\intercal)^\intercal$:
\begin{equation}
\label{eq:dcm}
\boldsymbol{C}(\boldsymbol{q})=
\left(q_0^2-\boldsymbol{q}_v^\intercal\boldsymbol{q}_v\right)\boldsymbol{I}_3
+2\,\boldsymbol{q}_v\boldsymbol{q}_v^\intercal
+2\,q_0\,[\boldsymbol{q}_v]_\times,
\end{equation}
where $[\boldsymbol{q}_v]_\times$ denotes the skew-symmetric cross-product matrix of $\boldsymbol{q}_v$.
\subsection{Geometry description}
For expressing constraints and pointing error, the unit direction vectors (from spacecraft) $\boldsymbol{v}_{\mathrm{trg}}^{\mathcal{N}}(t)$ (to target), $\boldsymbol{v}_{\mathrm{sun}}^{\mathcal{N}}(t)$ (to Sun), and $\boldsymbol{v}_{\mathrm{nadir}}^{\mathcal{N}}(t)$ (to nadir) are defined. These vectors evolve continuously as the spacecraft orbits and the observation geometry changes. A dedicated software module evaluates these vectors over the entire prediction horizon at each flight software step, using ephemeris data and astronomical models~\cite{cite:fundamentals}.

\subsubsection*{Pointing error}
The pointing error $\theta_{\mathrm{err}}$ is defined as the angle between the instrument boresight $\boldsymbol{v}_{\mathrm{ins}}^{\mathcal{B}}$ and the target direction rotated into body frame $\boldsymbol{v}_{\mathrm{trg}}^{\mathcal{B}}$:
\begin{equation}
\theta_{\mathrm{err}}(t) = \arccos\left(\left(\boldsymbol{v}_{\mathrm{ins}}^{\mathcal{B}}\right)^{\intercal}\bigg(\boldsymbol{C}(\boldsymbol{q}(t))\boldsymbol{v}_{\mathrm{trg}}^{\mathcal{N}}(t)\bigg)\right).
\label{eq:pointing_error}
\end{equation}

Because $\theta_{\mathrm{err}}$ depends only on the angle between two vectors, rotations around the boresight axis leave it unchanged. This residual degree of freedom can be used to satisfy the star-tracker exclusion constraints without degrading pointing accuracy.

\subsubsection*{Exclusion zone constraints}
To maintain star-tracker operability, two hard geometric exclusion constraints are enforced. The star-tracker boresight $\boldsymbol{v}_{\mathrm{str}}^{\mathcal{B}}$ must maintain a minimum angular separation from both the Sun and nadir directions:
\begin{equation}
\begin{aligned}
\left(\boldsymbol{v}_{\mathrm{str}}^{\mathcal{B}}\right)^{\intercal}\!\bigg(\boldsymbol{C}(\boldsymbol{q}(t))\boldsymbol{v}_{\mathrm{sun}}^{\mathcal{N}}(t)\bigg) &\le \cos\!\left(\alpha_{\mathrm{exc},\mathrm{sun}}\right),\\
\left(\boldsymbol{v}_{\mathrm{str}}^{\mathcal{B}}\right)^{\intercal}\!\bigg(\boldsymbol{C}(\boldsymbol{q}(t))\boldsymbol{v}_{\mathrm{nadir}}^{\mathcal{N}}(t)\bigg) &\le \cos\!\left(\alpha_{\mathrm{exc},\mathrm{nadir}}\right),
\end{aligned}
\end{equation}
for all $t$ during the maneuver, where $\alpha_{\mathrm{exc},\mathrm{sun}}$ and $\alpha_{\mathrm{exc},\mathrm{nadir}}$ are the exclusion half-angles.

\subsubsection*{Operational limits}
The satellite's rotational speed is constrained not to exceed a maximum angular rate in order to maintain stable star-tracker lock and attitude estimation accuracy:
\begin{equation}
\begin{aligned}
-\boldsymbol{\omega}_{\max} \le \boldsymbol{\omega}(t) \le \boldsymbol{\omega}_{\max}, \quad -\boldsymbol{u}_{\max} \le \boldsymbol{u}(t) \le \boldsymbol{u}_{\max},
\end{aligned}
\end{equation}
where, $\boldsymbol{\omega}_{\max}\,=\omega_{\max}\boldsymbol{1}_\mathrm{3\times 1}\,$ is selected according to star-tracker rate capability, and $u_{\max}$ is the per-axis actuator torque limit.

\section{Model predictive controller design}
\label{sec:mpc_design}

The nonlinear attitude dynamics and geometric constraint outputs from Section~II are analytically linearized at each sampling instant, resulting in a time-varying discrete linear model. A Kalman filter with augmented disturbance states provides state and disturbance estimates, and the MPC is implemented in a receding-horizon fashion~\cite{rawlings2020model}.

\subsection{Model of the Satellite}

\subsubsection{State dynamics}
The continuous-time dynamics are linearized about the operating point $\boldsymbol{x}_{\mathrm{lp}} = (\boldsymbol{0}^\intercal, \hat{\boldsymbol{q}}^\intercal)^\intercal$, where $\hat{\boldsymbol{q}}$ is the current quaternion estimate. At zero angular velocity, the continuous-time Jacobians reduce to
\begin{equation}
\label{eq:Ac}
\boldsymbol{A}_\text{c} =
\begin{pmatrix}
\boldsymbol{0}_{3\times 3} & \boldsymbol{0}_{3\times 4} \\[2pt]
\boldsymbol{A}_{\text{q}\omega}(\hat{\boldsymbol{q}}) & \boldsymbol{0}_{4\times 4}
\end{pmatrix}\!,\quad
\boldsymbol{B}_\text{c} =
\begin{pmatrix}
\boldsymbol{J}_{\mathcal{B}}^{-1} \\[2pt]
\boldsymbol{0}_{4\times 3}
\end{pmatrix}\!,
\end{equation}
with state dimension $n_x = 7$ and input dimension $n_u = 3$. The block $\boldsymbol{A}_{\text{q}\omega} \in \mathbb{R}^{4\times 3}$ is the quaternion-kinematics sensitivity
\begin{equation}
\label{eq:Aqw}
\boldsymbol{A}_{\text{q}\omega}(\hat{\boldsymbol{q}}) = \frac{1}{2}
\begin{pmatrix}
-\hat{q}_1 & -\hat{q}_2 & -\hat{q}_3 \\
\phantom{-}\hat{q}_0 & -\hat{q}_3 &  \phantom{-}\hat{q}_2 \\
 \phantom{-}\hat{q}_3 &  \phantom{-}\hat{q}_0 & -\hat{q}_1 \\
-\hat{q}_2 &  \phantom{-}\hat{q}_1 &  \phantom{-}\hat{q}_0
\end{pmatrix},
\end{equation}
derived from the quaternion multiplication matrix $\boldsymbol{Q}(\boldsymbol{q})$. The continuous-time pair $\boldsymbol{A}_\text{c},\boldsymbol{B}_\text{c}$ is discretized with exact zero-order hold at sampling period $T_s$ to obtain $\boldsymbol{A}_\text{d},\,\boldsymbol{B}_\text{d}$. The pair $\boldsymbol{A}_\text{d},\,\boldsymbol{B}_\text{d}$ is recomputed at every control step using the latest quaternion estimate.

\subsubsection{Geometric output linearization}
Using the DCM defined in~\eqref{eq:dcm}, the three scalar geometric outputs are
\begin{subequations}
\label{eq:outputs}
\begin{align}
y_{\mathrm{trg},k} &= \left(\boldsymbol{v}_{\mathrm{ins}}^{\mathcal{B}}\right)^{\!\intercal}\!\boldsymbol{C}(\boldsymbol{q}_k)\,\boldsymbol{v}_{\mathrm{trg}}^{\mathcal{N}}(k), \label{eq:y_trg}\\
y_{\mathrm{sun},k} &= \left(\boldsymbol{v}_{\mathrm{str}}^{\mathcal{B}}\right)^{\!\intercal}\!\boldsymbol{C}(\boldsymbol{q}_k)\,\boldsymbol{v}_{\mathrm{sun}}^{\mathcal{N}}(k), \label{eq:y_sun}\\
y_{\mathrm{nadir},k} &= \left(\boldsymbol{v}_{\mathrm{str}}^{\mathcal{B}}\right)^{\!\intercal}\!\boldsymbol{C}(\boldsymbol{q}_k)\,\boldsymbol{v}_{\mathrm{nadir}}^{\mathcal{N}}(k). \label{eq:y_nadir}
\end{align}
\end{subequations}

The first output~\eqref{eq:y_trg} drives the pointing objective, while~\eqref{eq:y_sun} and~\eqref{eq:y_nadir} enforce the star-tracker exclusion constraints. Each output depends nonlinearly on $\boldsymbol{q}$ through the direction cosine matrix and is linearized via a first-order Taylor expansion about $\boldsymbol{x}_{\mathrm{lp}}$:
\begin{equation}
\label{eq:y_lin}
y_k \approx \boldsymbol{h}_k^\intercal \boldsymbol{x}_k + c_k,
\end{equation}
where $\boldsymbol{h}_k \in \mathbb{R}^{n_x}$ is obtained from the analytical Jacobian and $c_k = \hat{y}_k - \boldsymbol{h}_k^\intercal\boldsymbol{x}_{\mathrm{lp}}$ is the affine offset, with $\hat{y}_k$ denoting the output evaluated by the nonlinear model at the current linearization point $\boldsymbol{x}_{\mathrm{lp}}$. Because~\eqref{eq:outputs} depends solely on $\boldsymbol{q}$, the entries of $\boldsymbol{h}_k$ corresponding to $\boldsymbol{\omega}$ are identically zero. The inertial direction vectors vary along the orbit and are assumed known over the prediction horizon from ephemeris propagation, yielding a distinct row $\boldsymbol{h}_k^\intercal$ at each step $k = 0,\ldots,N{-}1$, where $N$ denotes the prediction horizon length. 

\subsection{Design of the State and Disturbance Observer}
\label{sec:observer}

A time-varying Kalman filter (TVKF) is employed to estimate the state and disturbance channels under model-plant mismatch, similarly to~\cite{pannocchia2003disturbance}. The state $\boldsymbol{x} \in \mathbb{R}^{n_\text{x}}$ is augmented with two disturbance channels:
\begin{equation}
\label{eq:z_aug}
\boldsymbol{z} = \left(\boldsymbol{x}^\intercal,\,\boldsymbol{d}_\text{u}^\intercal,\,\boldsymbol{d}_\text{y}^\intercal\right)^\intercal \in \mathbb{R}^{n_\text{z}},\quad n_\text{z} = n_\text{x} + n_{\text{d}_\text{u}} + n_{\text{d}_\text{y}},
\end{equation}
where $\boldsymbol{d}_\text{u} \in \mathbb{R}^{n_{\text{d}_\text{u}}}$ with $n_{\text{d}_\text{u}} = n_\text{u} = 3$ models a persistent torque disturbance acting through the input channel, and $\boldsymbol{d}_\text{y} \in \mathbb{R}^{n_{\text{d}_\text{y}}}$ with $n_{\text{d}_\text{y}} = 3$ captures biases on the three geometric outputs~\eqref{eq:outputs}. Both disturbances follow a random-walk model. The augmented prediction step is
\begin{equation}
\label{eq:aug_pred}
\boldsymbol{z}_{k+1\mid k} = \boldsymbol{A}_{\mathrm{aug}}\,\boldsymbol{z}_{k\mid k} + \boldsymbol{B}_{\mathrm{aug}}\,\boldsymbol{u}_k + \boldsymbol{c}_{\mathrm{aff}},
\end{equation}
with
\begin{subequations}
\label{eq:aug_matrices}
\begin{align}
\boldsymbol{A}_{\mathrm{aug}} &=
\begin{pmatrix}
\boldsymbol{A}_\text{d} & \boldsymbol{B}_\text{d} & \boldsymbol{0} \\
\boldsymbol{0} & \boldsymbol{I}_{n_{\text{d}_\text{u}}} & \boldsymbol{0} \\
\boldsymbol{0} & \boldsymbol{0} & \boldsymbol{I}_{n_{\text{d}_\text{y}}}
\end{pmatrix}, \\
\boldsymbol{B}_{\mathrm{aug}} &=
\begin{pmatrix}
\boldsymbol{B}_\text{d} \\[1pt] \boldsymbol{0} \\[1pt] \boldsymbol{0}
\end{pmatrix}, \\
\boldsymbol{c}_{\mathrm{aff}} &=
\begin{pmatrix}
(\boldsymbol{I}-\boldsymbol{A}_\text{d})\boldsymbol{x}_{\mathrm{lp}} \\[1pt] \boldsymbol{0} \\[1pt] \boldsymbol{0}
\end{pmatrix}.
\end{align}
\end{subequations}

The input disturbance $\boldsymbol{d}_\text{u}$ enters through the $\boldsymbol{B}_\text{d}$ block, capturing unmodeled torques such as environmental disturbances and unmodeled actuator dynamics. The output disturbance $\boldsymbol{d}_\text{y}$ compensates for linearization errors and unmodeled dynamics in the geometric outputs.

The measurement vector $\boldsymbol{y}_k \in \mathbb{R}^{n_\text{x} + n_{\text{d}_\text{y}}}$ combines the full attitude state (from star-tracker quaternion and gyroscope rates) with the three geometric outputs in raw form:
\begin{subequations}
\label{eq:meas}
\begin{align}
\boldsymbol{y}_k &= \left(\boldsymbol{x}_k^\intercal,\; y_{\mathrm{trg},k},\; y_{\mathrm{sun},k},\; y_{\mathrm{nadir},k}\right)^\intercal, \\
\boldsymbol{y}_k &= \boldsymbol{C}_{\mathrm{aug},k}\boldsymbol{z}_k + \boldsymbol{b}_k,
\end{align}
\end{subequations}
where
\begin{equation}
\label{eq:meas_aff}
\boldsymbol{b}_k =
\left(\boldsymbol{0}_{n_x}^\intercal,\; c_{\mathrm{trg},0},\; c_{\mathrm{sun},0},\; c_{\mathrm{nadir},0}\right)^\intercal.
\end{equation}

The time-varying observation matrix $\boldsymbol{C}_{\mathrm{aug},k} \in \mathbb{R}^{(n_\text{x}+n_{\text{d}_\text{y}})\times n_\text{z}}$ is
\begin{equation}
\label{eq:Ck}
\boldsymbol{C}_{\mathrm{aug},k} =
\begin{pmatrix}
\boldsymbol{I}_{n_\text{x}} & \boldsymbol{0} & \boldsymbol{0} \\
\boldsymbol{h}_{\mathrm{trg},0}^\intercal & \boldsymbol{0}^\intercal & \boldsymbol{e}_1^\intercal \\
\boldsymbol{h}_{\mathrm{sun},0}^\intercal & \boldsymbol{0}^\intercal & \boldsymbol{e}_2^\intercal \\
\boldsymbol{h}_{\mathrm{nadir},0}^\intercal & \boldsymbol{0}^\intercal & \boldsymbol{e}_3^\intercal
\end{pmatrix}\!,
\end{equation}
where $\boldsymbol{e}_i$ denotes the $i$-th canonical basis vector of $\mathbb{R}^{n_{\text{d}_\text{y}}}$ and the subscript~$0$ indicates the current-step linearization. The upper block directly observes the state, while the lower rows relate the geometric outputs to the augmented state via the current linearization and include the output disturbance. The affine measurement term $\boldsymbol{b}_k$ carries the linearization offsets.


\subsection{MPC design}
\label{sec:mpc}

The controller is formulated in the standard receding-horizon form in deviation coordinates
$\delta\boldsymbol{x}_k = \boldsymbol{x}_k - \boldsymbol{x}_{\mathrm{lp}}$, with
$\boldsymbol{x}_{\mathrm{lp}} = (\boldsymbol{0}^\intercal,\hat{\boldsymbol{q}}^\intercal)^\intercal$.
The state ordering is $\boldsymbol{x}_k=(\boldsymbol{\omega}_k^\intercal,\boldsymbol{q}_k^\intercal)^\intercal$, i.e., angular rates first and quaternion states second.
The decision variable $\boldsymbol{U}=(\boldsymbol{u}_0^\intercal,\ldots,\boldsymbol{u}_{N-1}^\intercal)^\intercal\in\mathbb{R}^{Nn_u}$ stacks the body-torque inputs over the prediction horizon. The initial condition is set from the TVKF estimate at time $t$ as $\delta\boldsymbol{x}_0=\hat{\boldsymbol{x}}(t)-\boldsymbol{x}_{\mathrm{lp}}(t)$.
The optimization problem solved at each sampling instant is
\begin{subequations}
\label{eq:mpc}
\begin{align}
\min_{\boldsymbol{U},\,\boldsymbol{s}} \;\; & J(\boldsymbol{U},\,\boldsymbol{s}) \label{eq:mpc_obj} \\
\text{s.t.}\;\; & \delta\boldsymbol{x}_{k+1} = \boldsymbol{A}_\text{d}\,\delta\boldsymbol{x}_k + \boldsymbol{B}_\text{d}\,\boldsymbol{u}_k + \boldsymbol{B}_\text{d}\,\hat{\boldsymbol{d}}_u, \label{eq:mpc_dyn}\\
& y_{\mathrm{trg},k} = \boldsymbol{h}_{\mathrm{trg},k}^{\intercal}\delta\boldsymbol{x}_k + \bar{c}_{\mathrm{trg},k} + d_{y,1}, \label{eq:mpc_ytrg}\\
& y_{\mathrm{sun},k} = \boldsymbol{h}_{\mathrm{sun},k}^{\intercal}\delta\boldsymbol{x}_k + \bar{c}_{\mathrm{sun},k} + d_{y,2}, \label{eq:mpc_ysun}\\
& y_{\mathrm{nadir},k} = \boldsymbol{h}_{\mathrm{nadir},k}^{\intercal}\delta\boldsymbol{x}_k + \bar{c}_{\mathrm{nadir},k} + d_{y,3}, \label{eq:mpc_ynadir}\\
& -\boldsymbol{\omega}_{\max} \!-\! \boldsymbol{s}_{\omega,k} \leq \boldsymbol{\omega}_k \leq \boldsymbol{\omega}_{\max} \!+\! \boldsymbol{s}_{\omega,k}, \label{eq:mpc_omegabnd}\\
& y_{\mathrm{sun},k} \leq \cos\alpha_{\mathrm{sun}} + s_{\mathrm{sun},k}, \label{eq:mpc_sun}\\
& y_{\mathrm{nadir},k} \leq \cos\alpha_{\mathrm{nadir}} + s_{\mathrm{nadir},k}, \label{eq:mpc_nadir}\\
& -\boldsymbol{u}_{\max} \leq \boldsymbol{u}_k \leq \boldsymbol{u}_{\max}, \label{eq:mpc_ubnd}\\
& \boldsymbol{s} \geq \boldsymbol{0},\\
& \delta\boldsymbol{x}_0 = \hat{\boldsymbol{x}}(t) - \boldsymbol{x}_{\mathrm{lp}}(t),\\
& k = 0,\ldots,N{-}1, \label{eq:mpc_slack}
\end{align}
\end{subequations}
where $\boldsymbol{s} = \left(\boldsymbol{s}_{\omega}^\intercal, \boldsymbol{s}_{\mathrm{sun}}^\intercal, \boldsymbol{s}_{\mathrm{nadir}}^\intercal\right)^\intercal \in \mathbb{R}^{N(3+2)}$ collects slack variables introduced strictly as a feasibility safeguard: they prevent QP infeasibility when the nonlinear trajectory departs from the linearized admissible set. The high penalty $w_\text{s} = 10^9$ in~\eqref{eq:cost} ensures the original constraints are satisfied whenever the linearized feasible set is non-empty. 
The affine offsets are corrected for the shift by $\boldsymbol{x}_{\mathrm{lp}}$ as
$\bar{c}_{\mathrm{trg},k}=c_{\mathrm{trg},k}+\boldsymbol{h}_{\mathrm{trg},k}^{\intercal}\boldsymbol{x}_{\mathrm{lp}}$,
$\bar{c}_{\mathrm{sun},k}=c_{\mathrm{sun},k}+\boldsymbol{h}_{\mathrm{sun},k}^{\intercal}\boldsymbol{x}_{\mathrm{lp}}$, and
$\bar{c}_{\mathrm{nadir},k}=c_{\mathrm{nadir},k}+\boldsymbol{h}_{\mathrm{nadir},k}^{\intercal}\boldsymbol{x}_{\mathrm{lp}}$.

The cost function combines pointing accuracy, control regularization, and rate smoothness:
\begin{equation}
\label{eq:cost}
\begin{aligned}
J = \sum_{k=0}^{N-1}\Big( & w_\text{p}\!\left(y_{\mathrm{trg},k}-1\right)^2 + \boldsymbol{\omega}_k^\intercal\!\boldsymbol{Q}_\omega\,\boldsymbol{\omega}_k \\
& + \Delta\boldsymbol{\omega}_k^\intercal\!\boldsymbol{Q}_{\Delta\omega}\,\Delta\boldsymbol{\omega}_k + \Delta\boldsymbol{u}_k^\intercal\!\boldsymbol{Q}_{\Delta \text{u}}\,\Delta\boldsymbol{u}_k \\
& + w_\text{s}\!\left(\|\boldsymbol{s}_{\omega,k}\|^2 + s_{\mathrm{sun},k}^2 + s_{\mathrm{nadir},k}^2\right)\Big),
\end{aligned}
\end{equation}
where $w_\text{p}(y_{\mathrm{trg},k}-1)^2$ penalizes the squared deviation of the instrument-target cosine from unity (perfect alignment), $\boldsymbol{Q}_\omega$ penalizes angular velocity magnitude, $\boldsymbol{Q}_{\Delta\omega}$ penalizes angular-rate changes with $\Delta\boldsymbol{\omega}_0 = \boldsymbol{\omega}_0 - \boldsymbol{\omega}_{\mathrm{prev}}$ and $\Delta\boldsymbol{\omega}_k = \boldsymbol{\omega}_k - \boldsymbol{\omega}_{k-1}$ for $k \geq 1$, and $\boldsymbol{Q}_{\Delta \text{u}}$ suppresses input changes with $\Delta\boldsymbol{u}_0 = \boldsymbol{u}_0 - \boldsymbol{u}_{\mathrm{prev}}$ and $\Delta\boldsymbol{u}_k = \boldsymbol{u}_k - \boldsymbol{u}_{k-1}$ for $k \geq 1$. Here $\boldsymbol{\omega}_{\mathrm{prev}}$ and $\boldsymbol{u}_{\mathrm{prev}}$ denote the measured angular rate and the applied torque at the previous control step, respectively. All slack variables are penalized by the same high weight $w_\text{s}$ to strongly discourage boundary violations while keeping the optimization feasible.

For implementation, the problem~\eqref{eq:mpc} is condensed to a dense QP by eliminating the predicted states over the horizon,
\begin{equation}
\label{eq:qp}
\min_{\boldsymbol{p}}\;\tfrac{1}{2}\boldsymbol{p}^\intercal\!\boldsymbol{P}\,\boldsymbol{p} + \boldsymbol{g}^\intercal\!\boldsymbol{p}\quad\text{s.t.}\;\;\boldsymbol{A}_{\mathrm{ineq}}\,\boldsymbol{p} \leq \boldsymbol{b},\;\;\boldsymbol{p}_{\min} \leq \boldsymbol{p} \leq \boldsymbol{p}_{\max},
\end{equation}
with decision vector $\boldsymbol{p} = (\boldsymbol{U}^\intercal, \boldsymbol{s}_{\omega}^\intercal, \boldsymbol{s}_{\mathrm{sun}}^\intercal, \boldsymbol{s}_{\mathrm{nadir}}^\intercal)^\intercal \in \mathbb{R}^{N(n_u+3+2)}$. The Hessian $\boldsymbol{P}$ and linear term $\boldsymbol{g}$ are updated at every control step from the current TVKF estimates. The inequality matrix $\boldsymbol{A}_{\mathrm{ineq}}$ encodes the softened angular-rate bounds and exclusion constraints, while the box constraints $\boldsymbol{p}_{\min} \leq \boldsymbol{p} \leq \boldsymbol{p}_{\max}$ enforce actuator limits on $\boldsymbol{U}$ and non-negativity on all slacks. Only the first optimal input $\boldsymbol{u}_0^*$ is applied, and the procedure repeats at the next sampling instant.

\section{Nonlinear simulation}
\subsection{Simulation environment}
For the nonlinear simulations, NASA's "42" spacecraft high-fidelity dynamics simulator is employed~\cite{42simulator}. An 8U CubeSat with two flexible deployable solar panels is configured. The environmental disturbance torques for gravity gradient, aerodynamic, magnetic dipole, and radiation pressure torques are enabled. Ideal body torques are applied directly to the simulator without control allocation. The true simulated state is provided to the controller, i.e., no gyro, star-tracker, or attitude estimator is modeled. The observer described in Section~\ref{sec:observer} is part of the flight-software architecture but is not exercised in the present simulations, which use perfect state feedback. Observer validation under realistic sensor noise is left to future work. It is assumed that control torques in all axes can be directly commanded, generated by underlying actuator allocation. The only actuation parameter entering the MPC is the torque saturation bound $\boldsymbol{u}_{\max}$ for the given hardware configuration.

The guidance and control algorithms are implemented in Python as a prototype of the on-board flight software and are coupled with "42" in a model-in-the-loop configuration. Reference vectors for nadir, Sun, and target are generated from orbital data and ephemerides. The flight software is executed with a sampling period of $T_s = 0.1\,\mathrm{s}$, but the simulation step size is selected as $T_{sim} = 0.01\,\mathrm{s}$. The MPC at each control step is solved using the Gurobi quadratic programming (QP) solver.

\subsection{Simulation scenario description}
\label{subsec:sim_sc_desc}
An 8U Earth-observation CubeSat in a $550\,\mathrm{km}$ sun-synchronous orbit (SSO) is considered. The key parameters of the nonlinear simulation scenario are summarized in Table~\ref{tab:spacecraft_parameters}. The payload instrument has a field of view (FOV) corresponding to a $20\,\mathrm{km}$ wide ground swath. With a maximum pointing-error tolerance of $1^{\circ}$, proper image alignment and data quality are ensured. For targets up to $30^{\circ}$ off-nadir, a $1^{\circ}$ pointing error displaces the imaged footprint by at most $11\,\mathrm{km}$, so the target remains within the instrument FOV. The off-nadir angle is defined as the angle between the target direction $\boldsymbol{v}_{\mathrm{trg}}^{\mathcal{N}}$ and the nadir direction $\boldsymbol{v}_{\mathrm{nadir}}^{\mathcal{N}}$.

\begin{table}[!httb]
    \centering
    \caption{Spacecraft parameters}
    \label{tab:spacecraft_parameters}
    \renewcommand{\arraystretch}{1.2}
    \begin{tabular}{lll}
        \hline
        Symbol & Description & Value \\
        \hline
        $\boldsymbol{J}_{\mathcal{B}}$ & Inertia tensor in body frame & see~\eqref{eq:J_B} \\
        $\boldsymbol{v}_{\mathrm{ins}}^{\mathcal{B}}$ & Instrument boresight direction & $(0,\,0,\,1)^{\intercal}$ \\
        $\boldsymbol{v}_{\mathrm{str}}^{\mathcal{B}}$ & Star-tracker boresight direction & $(0,\,0.97,\,-0.23)^{\intercal}$ \\
        $\alpha_{\mathrm{exc},\mathrm{sun}}$ & Sun exclusion half-angle & $45^\circ$ \\
        $\alpha_{\mathrm{exc},\mathrm{nadir}}$ & Nadir exclusion half-angle & $89^\circ$ \\
        $\omega_{\max}$ & Maximum angular rate(per axis) & $3\si{^\circ\,s^{-1}}$ \\
        $u_{\max}$ & Maximum body torque (per axis) & $0.002\,\mathrm{N\,m}$ \\
        \hline
    \end{tabular}
    \renewcommand{\arraystretch}{1.0}
\end{table}

The nonlinear simulation run is $200\,\mathrm{s}$ long. Each simulation is initialized in the same nadir-pointing attitude that is selected such that the star-tracker boresight avoids exclusion zones posed by the Sun and nadir. The spacecraft performs a slew maneuver towards the line-of-sight direction of a fixed ground location corresponding to Prague and then tracks this target until the end of the simulation. The time of closest approach to the target is $t = 100\,\mathrm{s}$, at which the off-nadir angle is $26.7^{\circ}$.

The numerical inertia tensor used in the simulations is
\begin{equation}
\label{eq:J_B}
    \boldsymbol{J}_{\mathcal{B}} =
    \begin{pmatrix}
        \phantom{-}0.1335 & -0.0015 & \phantom{-}0.0045 \\
        -0.0015 & \phantom{-}0.1545 & -0.0225 \\
        \phantom{-}0.0045 & -0.0225 & \phantom{-}0.1065
    \end{pmatrix} \si{kg\,m^2}.
\end{equation}

\subsection{Guidance and control approaches}
\subsubsection{Trajectory optimization}
The first benchmark approach is obtained by solving a nonlinear optimal control problem (OCP) for the full rigid-body attitude dynamics. The state and input follow the notation of Section~\ref{sec:problem_statement}. Over the maneuver horizon $t \in [0,T_f]$ with $T_f = 200\,\mathrm{s}$, the OCP reads
\begin{subequations}
\label{eq:ocp_nonlinear}
\begin{align}
    \min_{\boldsymbol{\omega}(\cdot),\,\boldsymbol{q}(\cdot),\,\boldsymbol{u}(\cdot)} \; & \int_{0}^{T_f} \Big( - (\boldsymbol{v}_{\mathrm{ins}}^{\mathcal{B}})^{\intercal}\boldsymbol{C}(\boldsymbol{q}(t))\,\boldsymbol{v}_{\mathrm{trg}}^{\mathcal{N}}(t) \\
    & \hspace{6.0em} + \boldsymbol{u}(t)^{\intercal}\boldsymbol{R}\boldsymbol{u}(t) \Big)\,\mathrm{d}t \\[2pt]
    \text{s.t.}\quad & \dot{\boldsymbol{\omega}}(t)=\boldsymbol{J}_{\mathcal{B}}^{-1}\left(\boldsymbol{u}(t)-\boldsymbol{\omega}(t)\times\left(\boldsymbol{J}_{\mathcal{B}}\boldsymbol{\omega}(t)\right)\right), \\
    & \dot{\boldsymbol{q}}(t)=\frac{1}{2}\boldsymbol{Q}(\boldsymbol{q}(t))\begin{pmatrix}
    0 \\
    \boldsymbol{\omega}(t)
    \end{pmatrix}, \\
    & \|\boldsymbol{q}(t)\|^2 = 1, \\
    & \angle\big(\boldsymbol{v}_{\mathrm{str}}^{\mathcal{B}},\,\boldsymbol{C}(\boldsymbol{q}(t))\boldsymbol{v}_{\mathrm{sun}}^{\mathcal{N}}(t)\big) \geq \alpha_{\mathrm{exc},\mathrm{sun}}, \\
    & \angle\big(\boldsymbol{v}_{\mathrm{str}}^{\mathcal{B}},\,\boldsymbol{C}(\boldsymbol{q}(t))\boldsymbol{v}_{\mathrm{nadir}}^{\mathcal{N}}(t)\big) \geq \alpha_{\mathrm{exc},\mathrm{nadir}}, \\
    & -\boldsymbol{\omega}_{\max} \leq \boldsymbol{\omega}(t) \leq \boldsymbol{\omega}_{\max}, \\
    & -\boldsymbol{u}_{\max} \leq \boldsymbol{u}(t) \leq \boldsymbol{u}_{\max}, \\
    & \boldsymbol{x}(0) = \boldsymbol{x}_0,
\end{align}
\end{subequations}
where $\boldsymbol{R}\,=\,10^4\,\boldsymbol{I}_3$ is control-effort weighting matrix. The geometric quantities and system dynamics are as defined in Section~\ref{sec:problem_statement}.

The OCP~\eqref{eq:ocp_nonlinear} is transcribed into a nonlinear program (NLP) using a pseudospectral method on a Chebyshev--Gauss--Lobatto (CGL) grid with 40 points. The NLP is solved offline, and the resulting optimal state and input trajectories are represented by CGL polynomial in time~\cite{cite:pseudospectral}.

During nonlinear simulation, the on-board flight software evaluates these polynomials every $T_s = 0.1\,\mathrm{s}$. The feed-forward body torque $\boldsymbol{u}_{ff}(t)$, reference quaternion $\boldsymbol{q}(t)$ and rotational speed $\boldsymbol{\omega}(t)$ for the inner loop are computed. The inner loop is a state-feedback controller~\cite{cite:yang_quat}.

\subsubsection{Linear MPC}
An MPC described in Section~\ref{sec:mpc} was discretized with period $T_s = 0.1\,\mathrm{s}$. The prediction horizon was selected as $N = 50$ steps. The horizon choice is tightly linked to pointing performance. Increasing $N$ increases linearization error over the horizon, thus increasing pointing error.  A too-short horizon leads to a more aggressive input profile and an oscillatory pointing response. The objective-function weights used in this nonlinear simulation campaign are summarized in Table~\ref{tab:linear_mpc_obj_weights}.

\begin{table}[!httb]
    \centering
    \caption{Linear MPC objective-function weights used in simulation (Section~III-C).}
    \label{tab:linear_mpc_obj_weights}
    \renewcommand{\arraystretch}{1.15}
    \begin{tabular}{lll}
        \hline
        Term in \eqref{eq:cost} & Symbol & Value \\
        \hline
        Pointing term weight & $w_p$ & $100$ \\
        Angular-rate weight & $\boldsymbol{Q}_\omega$ & $0.05\,\boldsymbol{I}_3$ \\
        Rate-difference weight & $\boldsymbol{Q}_{\Delta\omega}$ & $\boldsymbol{I}_3$ \\
        Input-difference weight & $\boldsymbol{Q}_{\Delta u}$ & $\boldsymbol{I}_3$ \\
        Slack weight (all slacks) & $w_s$ & $10^9$ \\
        \hline
    \end{tabular}
    \renewcommand{\arraystretch}{1.0}
\end{table}
\subsubsection{Naive solution}
The third controller, referred to as the naive solution, is based on the quaternion reference-shaper approach from~\cite{cite:reference_governor} without any optimization-based governor. The same inner-loop state-feedback attitude controller as in the previous subsection is used, but its references are pre-shaped to respect only the angular-rate limits. At each flight-software step, the reference quaternion is computed using the TRIAD method~\cite{cite:fundamentals}, aligning the instrument boresight $\boldsymbol{v}_{\mathrm{ins}}^{\mathcal{B}}$ with the target direction $\boldsymbol{v}_{\mathrm{trg}}^{\mathcal{N}}$ and the star-tracker boresight $\boldsymbol{v}_{\mathrm{str}}^{\mathcal{B}}$ approximately opposite to the Sun direction $\boldsymbol{v}_{\mathrm{sun}}^{\mathcal{N}}$.

At each sampling instant $k$, an unconstrained quaternion reference $\boldsymbol{q}_{\mathrm{ref},k}$ is obtained from the ideal attitude that points the instrument boresight $\boldsymbol{v}_{\mathrm{ins}}^{\mathcal{B}}$ towards the ground-target direction $\boldsymbol{v}_{\mathrm{trg}}^{\mathcal{N}}(t_k)$, ignoring exclusion constraints. A rate-limited reference quaternion $\boldsymbol{q}_{\mathrm{ref},k}^{\mathrm{lim}}$ is then generated by spherical linear interpolation (SLERP) between the previously applied limited reference $\boldsymbol{q}_{\mathrm{ref},k-1}^{\mathrm{lim}}$ and the new unconstrained reference $\boldsymbol{q}_{\mathrm{ref},k}$,
\begin{equation}
    \boldsymbol{q}_{\mathrm{ref},k}^{\mathrm{lim}} = \mathrm{slerp}\big(\boldsymbol{q}_{\mathrm{ref},k-1}^{\mathrm{lim}}, \, \boldsymbol{q}_{\mathrm{ref},k}, \, \gamma_k\big),
\end{equation}
where the interpolation factor is chosen as $\gamma_k = \min(\alpha_{\max}/\alpha_{\mathrm{act}},\,1)$. Here $\alpha_{\max} = \omega_{\max} T_s$ is the maximum admissible rotation per sampling period and $\alpha_{\mathrm{act}}$ is the rotation angle between $\boldsymbol{q}_{\mathrm{ref},k-1}^{\mathrm{lim}}$ and $\boldsymbol{q}_{\mathrm{ref},k}$. The corresponding rate reference $\boldsymbol{\omega}_{\mathrm{ref},k}$ is obtained from the finite difference of the rate-limited quaternion sequence and used together with $\boldsymbol{q}_{\mathrm{ref},k}^{\mathrm{lim}}$ in the inner-loop error computation.

This naive controller enforces the angular-rate constraint through reference shaping alone. It does not explicitly account for Sun and nadir exclusion constraints or pointing-error tolerances, providing a baseline for comparison with the proposed MPC-based approach.

\begin{figure}[t]
    \centering
    \includegraphics[width=0.75\linewidth]{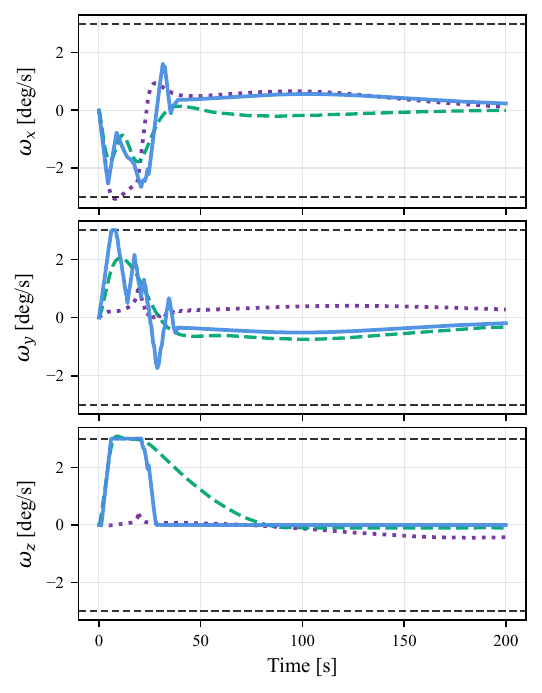}
    \caption{Angular-rate trajectories with rate limits. \textcolor[HTML]{4A90E2}{Solid blue}: Linear MPC. \textcolor[HTML]{00A676}{Dashed green}: nonlinear optimization. \textcolor[HTML]{7030A0}{Dotted purple}: naive solution. Black dashed lines: rate constraints.}
    \label{fig:comparison_control_omega_bounds}
\end{figure}

\begin{figure}[t]
    \centering
    \includegraphics[width=0.75\linewidth]{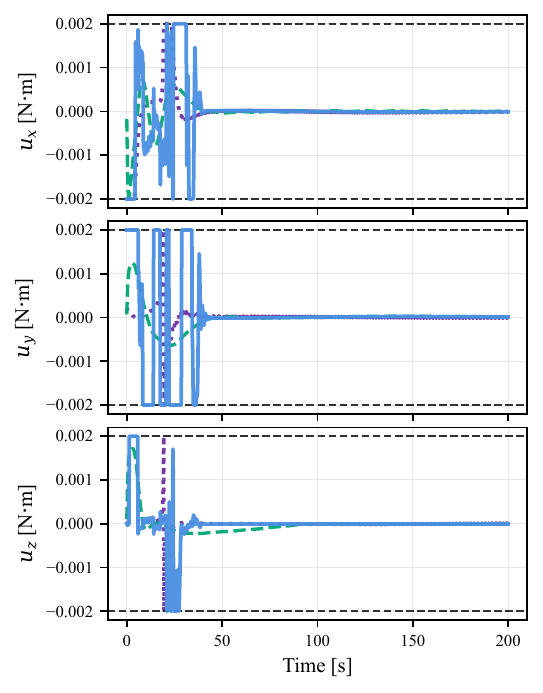}
    \caption{Body torque trajectories with actuator limits. \textcolor[HTML]{4A90E2}{Solid blue}: Linear MPC. \textcolor[HTML]{00A676}{Dashed green}: nonlinear optimization. \textcolor[HTML]{7030A0}{Dotted purple}: naive solution. Black dashed lines: actuator constraints.}
    \label{fig:comparison_control_torque_bounds}
\end{figure}

\subsection{Single run results}
\label{subsec:single_run_results}

\begin{figure*}[t]
    \centering
    \subfigure[Instrument boresight at $t = 200\,\si{s}$\label{fig:comparison_instr_vector}]{\includegraphics[width=0.23\textwidth]{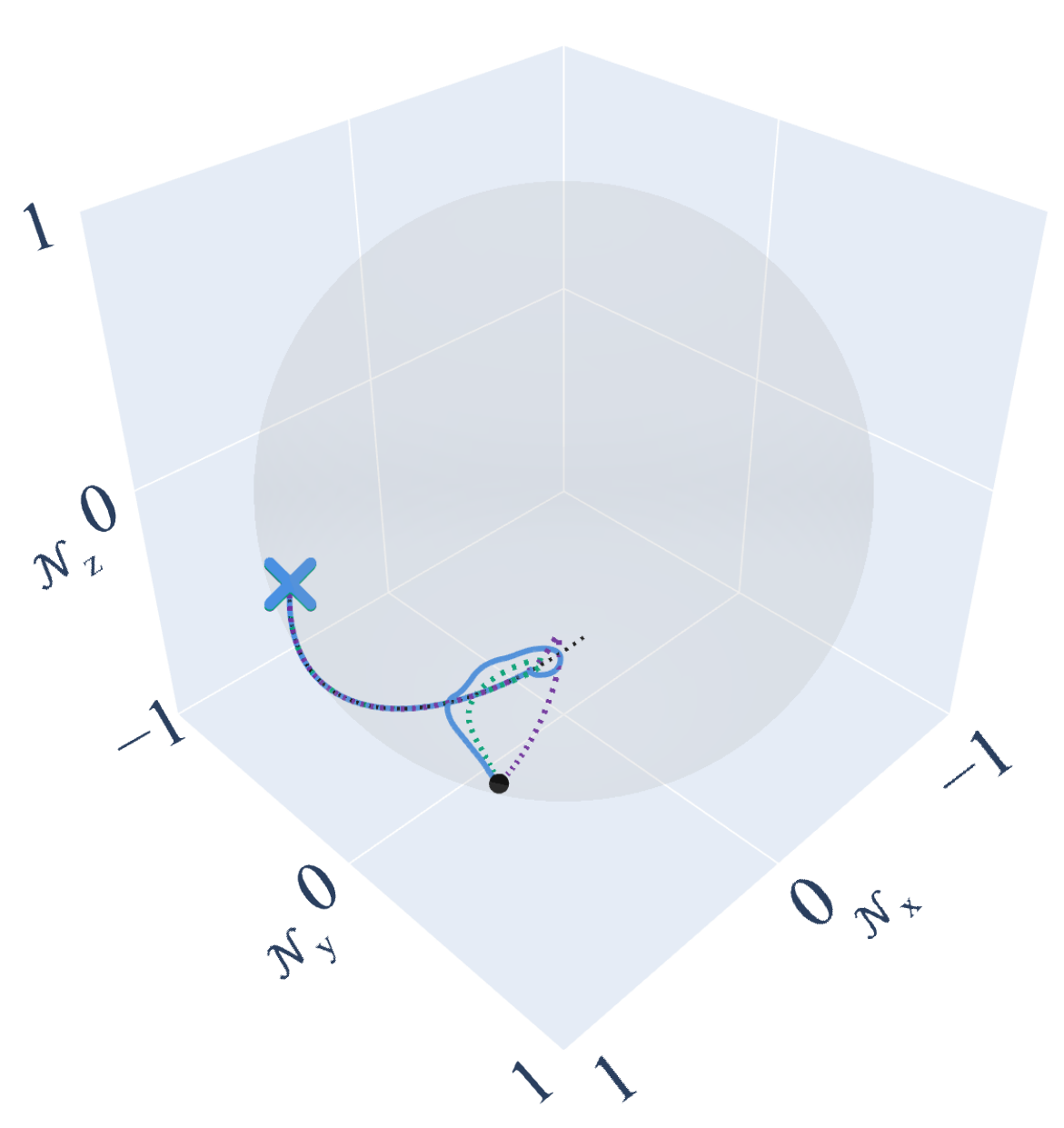}}\hfill
    \subfigure[Star-tracker boresight at $t = 10\,\si{s}$\label{fig:comparison_str_vector10}]{\includegraphics[width=0.23\textwidth]{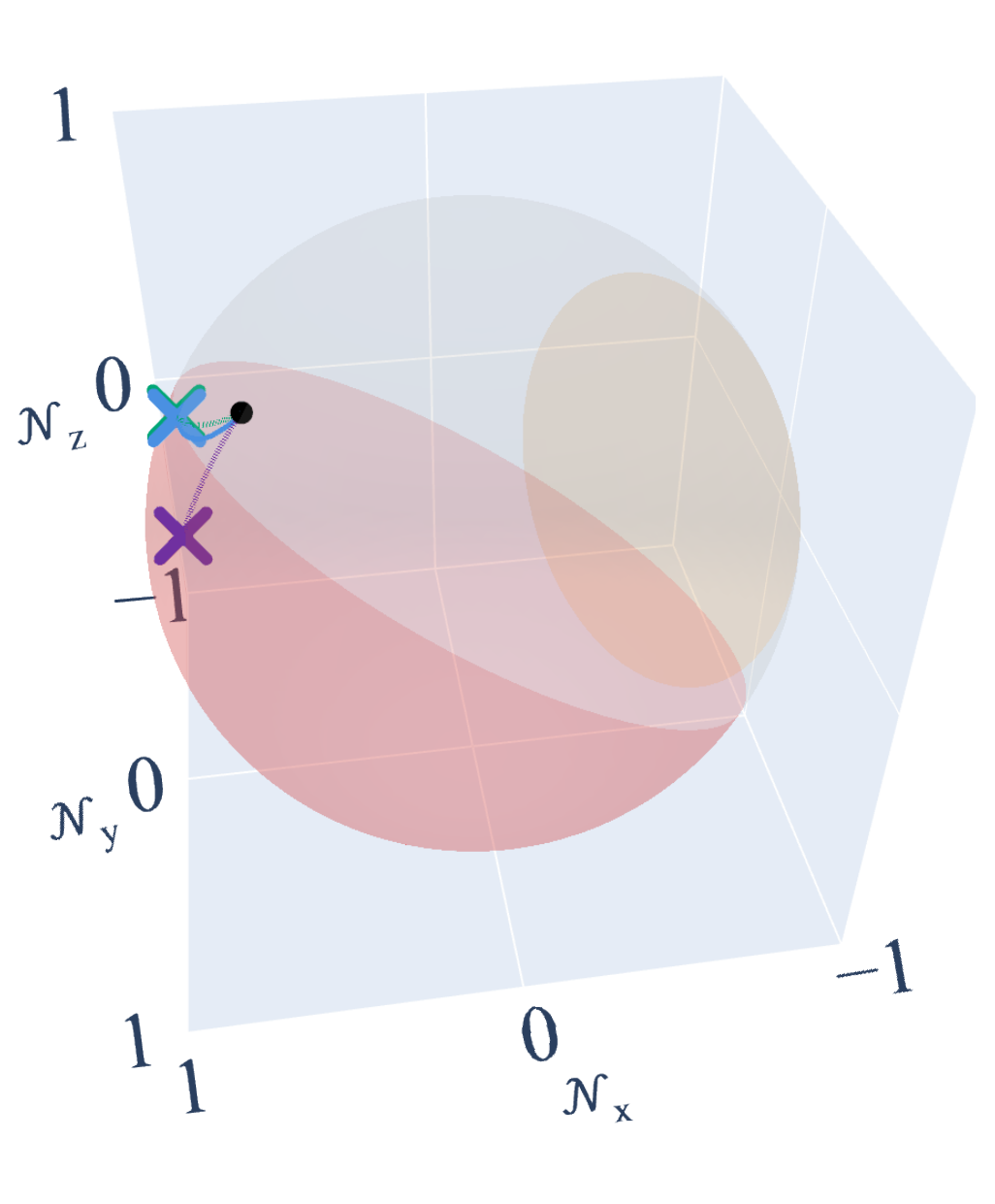}}\hfill
    \subfigure[Star-tracker boresight at $t = 28\,\si{s}$\label{fig:comparison_str_vector28}]{\includegraphics[width=0.23\textwidth]{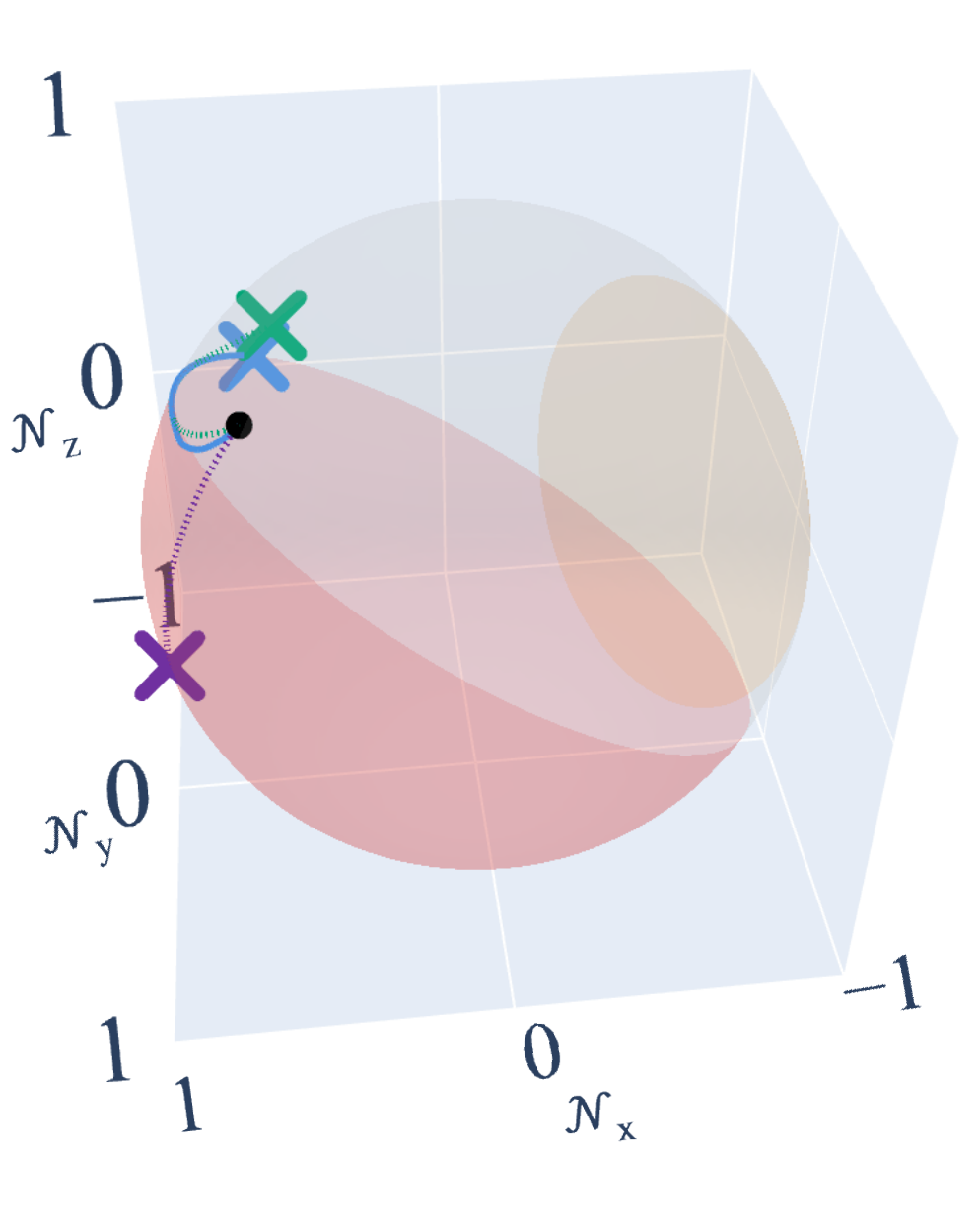}}\hfill
    \subfigure[Star-tracker boresight at $t = 200\,\si{s}$\label{fig:comparison_str_vector200}]{\includegraphics[width=0.23\textwidth]{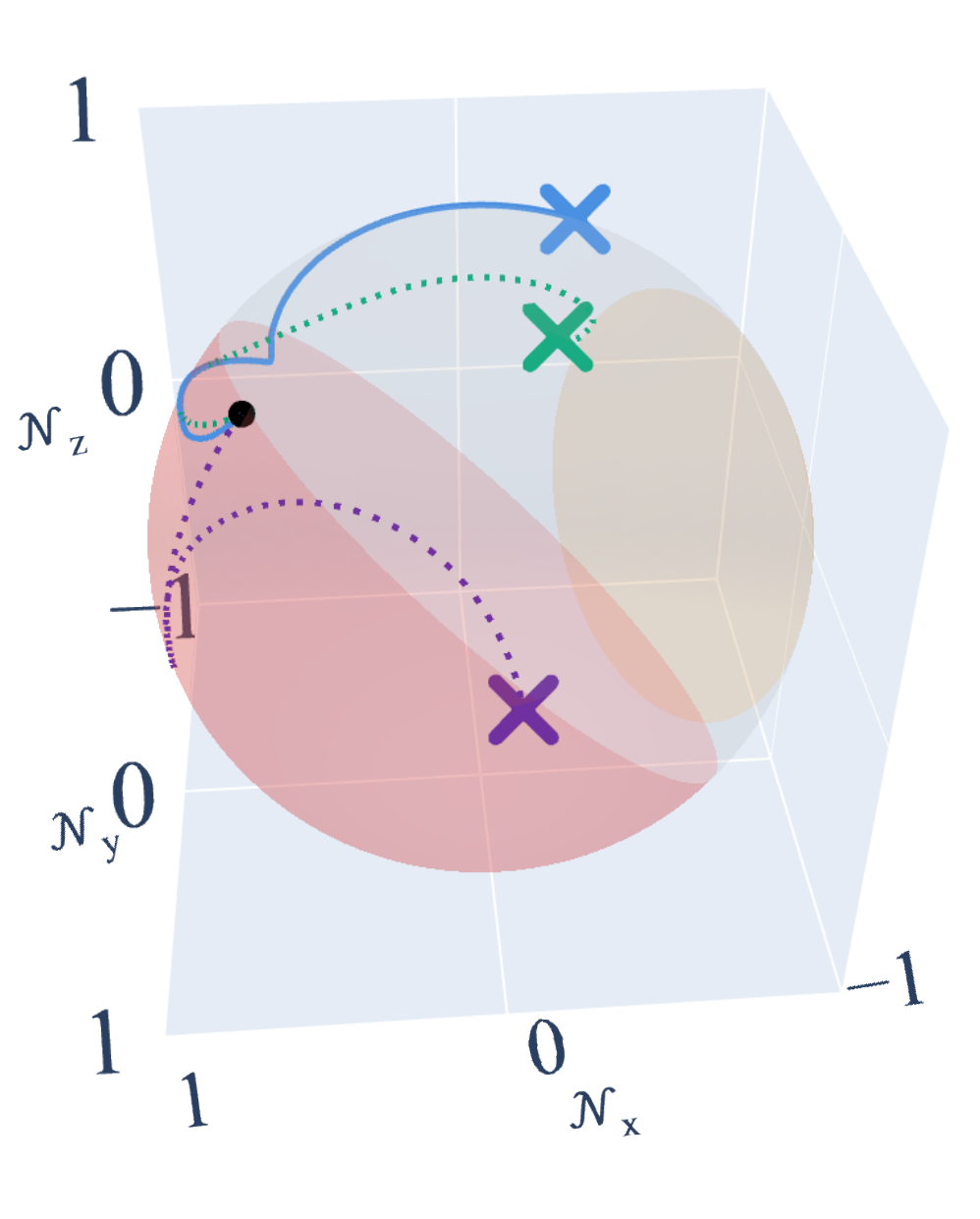}}
    \caption{Instrument $\boldsymbol{v}_\mathrm{ins}^\mathcal{N}$  and star-tracker boresight $\boldsymbol{v}_\mathrm{str}^\mathcal{N}$ (expressed in the inertial frame $\mathcal{N}$) visualization during the maneuver. Black point represents the vector direction at the maneuver start, and the cross represents the star-tracker boresight at time $t$. The black dashed line stands for $\boldsymbol{v}_\mathrm{trg}^\mathcal{N}$. The red set represents the nadir exclusion zone, and the yellow set represents the sun exclusion zone. \textcolor[HTML]{4A90E2}{Solid blue}: Linear MPC. \textcolor[HTML]{00A676}{Dashed green}: nonlinear optimization. \textcolor[HTML]{7030A0}{Dotted purple}: naive solution.}
    \label{fig:comparison_sphere}
\end{figure*}

The controllers described in the previous section are simulated in the scenario defined in Section~\ref{subsec:sim_sc_desc}. Figures~\ref{fig:comparison_control_omega_bounds} and~\ref{fig:comparison_control_torque_bounds} show the angular-rate and torque trajectories for all three approaches. Figure~\ref{fig:comparison_sphere} shows the instrument and star-tracker boresight directions projected into the inertial frame. Figures~\ref{fig:comparison_str_vector10},~\ref{fig:comparison_str_vector28}, and~\ref{fig:comparison_str_vector200} show the star-tracker boresight evolution relative to the time-varying Sun and nadir exclusion cones. After the transient, the target direction $\boldsymbol{v}_\mathrm{trg}^\mathcal{N}$ is tracked by all controllers.

Trajectory optimization exploits the full maneuver horizon, producing smooth control profiles with both exclusion zone constraints active. The linear MPC exhibits a more aggressive input profile due to the shorter prediction horizon. The Sun exclusion zone was not active for this scenario. The larger pointing error of the MPC relative to the trajectory optimization is attributed to linearization errors over the prediction horizon. The naive solution reaches the target fastest during the transient but ignores the exclusion zones entirely. Due to inner-loop controller overshoot, the rotational speed constraint is violated by approximately $3\%$. Quantitative single-run performance metrics after the settling is reached are summarized in Table~\ref{tab:single_run_results}. Settling time is defined as the first time instant at which the pointing error falls and remains below $1^\circ$ for a continuous window of three seconds. All three approaches tracked the target within the required $1^\circ$ accuracy, although each followed a different star-tracker boresight trajectory.

\begin{table}[!httb]
    \centering
    \caption{Single run performance metrics}
    \label{tab:single_run_results}
    \renewcommand{\arraystretch}{1.15}
    \begin{tabular}{lccc}
        \hline
        Metric & Linear MPC & Trajectory opt. & Naive \\
        \hline
        Mean pointing error [$^\circ$] & 0.188 & 0.031 & 0.035 \\
        Max pointing error [$^\circ$] & 0.412 & 0.110 & 0.112 \\
        Settling time [s] & 49.7 & 53.7 & 45.4 \\
        \hline
    \end{tabular}
    \renewcommand{\arraystretch}{1.0}

\end{table}

\begin{figure}[t]
    \centering
    \includegraphics[width=0.75\linewidth]{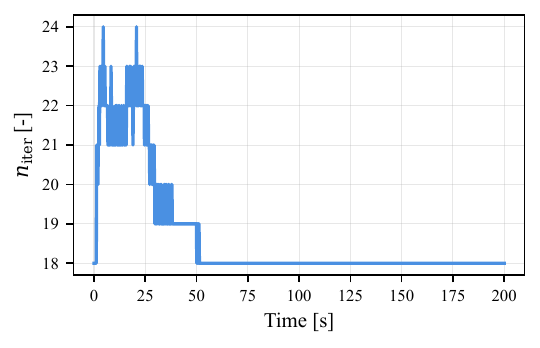}
    \caption{Number of QP solver iterations at each control step for the linear MPC controller.}
    \label{fig:mpc_timing_iter}
\end{figure}

For the linear MPC controller, the number of QP iterations can be seen in Figure~\ref{fig:mpc_timing_iter}. For the trajectory optimization approach, the algorithm took 89 iterations of the NLP solver.


\subsection{Monte Carlo (MC) simulations}

To assess the robustness of the proposed approach with respect to target geometry and parametric uncertainty of $\boldsymbol{J}_{\mathcal{B}}$, an MC campaign is carried out. The nonlinear simulation setup is the same as described in Section~\ref{subsec:sim_sc_desc}. The MC campaign is run only for Linear MPC, with the unchanged configuration.

For each run, a ground target is sampled on the Earth's surface by drawing latitude and longitude so that a wide range of geometries is excited. Only those sampled observations that satisfy two geometric validity conditions are retained: (i) the off-nadir angle between the instrument boresight and the local nadir direction at closest approach is below $30^{\circ}$, and (ii) the observation does not occur during the night. These conditions ensure Sun and nadir exclusion zones for the star-tracker remain applicable.


For each MC realization, the moments and products of inertia are perturbed by independent relative variations of up to $\pm30\,\%$. The perturbed positive definite $\boldsymbol{J}_{\mathcal{B}}$ is used for "42" simulator initialization.

The exclusion constraints were active in 97\% of the MC runs. In all of them, the exclusion zone, rotational speed, and input constraints were met.

Analysis of the pointing error reveals that in 11 out of 100 MC runs, both exclusion zone constraints were simultaneously active, causing the instrument boresight to drift away from the target. In these 11 runs, the star-tracker exclusion constraints remained satisfied, but the competing geometric demands caused the pointing error to exceed the $1^{\circ}$ tolerance. These runs represent a known geometric limitation and are excluded from the pointing-error analysis. In the remaining 89 runs, the mean pointing error during the observation (after transient) remained below $1^\circ$ in every case. In 65 of these runs, the pointing error stayed below $1^\circ$ throughout the entire observation window. In the remaining 24 runs, the pointing error temporarily exceeded the threshold near closest approach. The root cause is the rapid change in target, Sun, and nadir directions in the body frame. It introduces linearization errors over the prediction horizon that transiently degrade pointing accuracy. The results from the MC campaign are shown in the Table \ref{tab:mc_results}.

\begin{table}[!httb]
    \centering
    \caption{Monte Carlo performance metrics for the linear MPC approach over 89 runs with no simultaneous exclusion  zone constraints activation.}
    \label{tab:mc_results}
    \begin{tabular}{lll}
        \hline
        Quantity & Mean & Maximum \\ \hline
        Solver iterations$[-]$ & 19.28 & 29 \\
        Pointing error$[^\circ]$ & 0.31 & 2.95 \\
        Settling time$[\si{s}]$ & 44.0 & 72.5 \\
        \hline
    \end{tabular}
\end{table}

\section{Conclusion}

This paper presented a time-varying linear MPC framework for agile ground-target tracking on CubeSats with a single star-tracker. Analytical linearization of the rigid-body dynamics and geometric constraint outputs about the current state estimate at each control step reduces the online problem to a QP.

High-fidelity model-in-the-loop simulations using NASA's "42" simulator confirmed that the controller jointly satisfies angular-rate, actuator, and star-tracker exclusion constraints, without precomputed trajectories or ground intervention. In a representative single-run scenario, the MPC achieved sub-$1^\circ$ mean pointing accuracy, comparable to an offline nonlinear trajectory optimization while providing online feedback capability that the naive rate-limited baseline lacks. A Monte Carlo campaign over 100 randomized target geometries and inertia perturbations of up to 30\% demonstrated robust constraint satisfaction: exclusion constraints were active in 97\% of runs and satisfied in all 100. In 11 out of 100 runs, simultaneous activation of both exclusion zones caused sustained pointing drift — a characterized geometric limitation of the single-star-tracker configuration. Of the remaining 89 runs, 65 maintained pointing error below $1^\circ$ throughout the observation, while 24 exhibited transient violations near closest approach due to rapid geometry changes. The mean QP iteration count of approximately 19 per control step confirms the computational regularity of the approach.

Future work will focus on closing the gap between the presented design and full flight implementation, including integration of realistic gyroscope, star-tracker, and reaction wheel models to validate the observer under hardware-representative conditions. Strategies for detecting and handling geometrically infeasible configurations near closest approach will also be investigated. Formal recursive feasibility analysis and timing certification on flight-representative processors remain open problems toward mission deployment.

\section*{Acknowledgment}
D. Beňo is supported by the Grant Agency of the Czech Technical University in Prague, grant No. SGS25/145/OHK3/3T/13.
M. Klaučo is supported by the European Union project ROBOPROX (Reg. No. CZ.02.01.01/00/22\_008/0004590). 

P. Valábek gratefully acknowledges the contribution of the Scientific Grant Agency of the Slovak Republic under the grants VEGA 1/0239/24 and is supported by an internal STU grant for teams of young researchers and acknowledges the contribution of the EIT Manufacturing – Slovakia – X Fund.

\bibliography{references}{}
\bibliographystyle{unsrt}

\end{document}